\documentclass[letterpaper,aps,twocolumn,showpacs,superscriptaddress,floats,prl]{revtex4}

\usepackage[T1]{fontenc}
\usepackage[utf8]{inputenc}
\usepackage{graphicx,color}
\usepackage{amsmath}
\usepackage{wasysym,latexsym}

\usepackage{amssymb,dsfont}
\usepackage{textcomp}
\usepackage{booktabs}

\pacs{75.30.Ds, 78.70.Nx, 71.27.+a, 71.70.Gm}

\begin{document}
\title{
Band structure of helimagnons in MnSi resolved by inelastic neutron scattering}
\author{M. Kugler}
\email[e-mail: ]{mkugler@frm2.tum.de}
\author{G. Brandl}
\affiliation{Physik-Department E21, Technische Universit\"at M\"unchen, 85748 
Garching, Germany}
\affiliation{Heinz Maier-Leibnitz Zentrum (MLZ), Technische Universit\"at 
M\"unchen, 85748 Garching, Germany}
\author{J. Waizner}
\affiliation{Institut f\"ur Theoretische Physik, Universit\"at zu K\"oln, Z\"ulpicher Str. 77a, 50937 K\"oln, Germany}
\author{M. Janoschek}
\affiliation{Condensed Matter and Magnet Science, Los Alamos National 
Laboratory, Los Alamos, New Mexico 87545, USA}
\author{R. Georgii}
\affiliation{Physik-Department E21, Technische Universit\"at M\"unchen, 85748 
Garching, Germany}
\affiliation{Heinz Maier-Leibnitz Zentrum (MLZ), Technische Universit\"at 
M\"unchen, 85748 Garching, Germany}
\author{A.~Bauer}
\affiliation{Physik-Department E21, Technische Universit\"at M\"unchen, 85748 
Garching, Germany}
\author{K. Seemann}
\affiliation{Physik-Department E21, Technische Universit\"at M\"unchen, 85748 
Garching, Germany}
\affiliation{Heinz Maier-Leibnitz Zentrum (MLZ), Technische Universit\"at 
M\"unchen, 85748 Garching, Germany}
\author{A. Rosch}
\affiliation{Institut f\"ur Theoretische Physik, Universit\"at zu K\"oln, Z\"ulpicher Str. 77a, 50937 K\"oln, Germany}
\author{C. Pfleiderer}
\author{P. B\"oni}
\affiliation{Physik-Department E21, Technische Universit\"at M\"unchen, 85748 
Garching, Germany}
\author{M. Garst}
\affiliation{Institut f\"ur Theoretische Physik, Universit\"at zu K\"oln, Z\"ulpicher Str. 77a, 50937 K\"oln, Germany}
\date{\today}

\begin{abstract}

A magnetic helix realizes a one-dimensional magnetic crystal with a period given by the pitch length $\lambda_h$. 
Its spin-wave excitations -- the helimagnons -- experience Bragg scattering off this periodicity leading to gaps in the spectrum that inhibit their propagation along the pitch direction. Using high-resolution inelastic neutron scattering the resulting band structure of helimagnons was resolved by preparing a single crystal of MnSi in a single magnetic-helix domain. At least five helimagnon bands could be identified that cover the crossover from flat bands at low energies with helimagnons basically localized along the pitch direction to dispersing bands at higher energies. In the low-energy limit, we find the helimagnon spectrum to be determined by a universal, parameter-free theory. Taking into account corrections to this low-energy theory, quantitative agreement is obtained in the entire energy range studied with the help of a single fitting parameter.
\end{abstract}

\maketitle

The weak spin-orbit Dzyaloshinskii-Moriya interaction, $D$, in the cubic chiral magnets energetically favours spatial modulations of the magnetization. This gives rise to magnetic crystalline phases with unit cells that are incommensurate with and much larger than the atomic lattice spacing. Most prevalent is the magnetic helix, a one-dimensional magnetic crystal, with a large pitch $\lambda_h = 2\pi/k_h$ proportional to the ratio $J/D$ where $J$ is the magnetic exchange \cite{Bak_Jensen}. For a small range of finite magnetic fields, a two-dimensional magnetic crystal is also stabilized close to the critical temperature \cite{muehlbauer2009}. It can be identified as a lattice of magnetic skyrmions whose non-trivial topology is at the origin of various interesting phenomena \cite{Nagaosa2013} like, for example, a topological Hall effect \cite{Neubauer2009,Franz2014} and an emergent electrodynamics \cite{Jonietz2010,Schulz2012}. Interestingly, the phase transition from the paramagnetic to the magnetically ordered phases at small fields corresponds to a weak crystallization process \cite{Brazovskii1987} and is driven first-order by strongly correlated chiral paramagnons \cite{Brazowskii,Bauer2013,Buhrandt2013}.

\begin{figure*}[t]
\includegraphics[width=\linewidth] {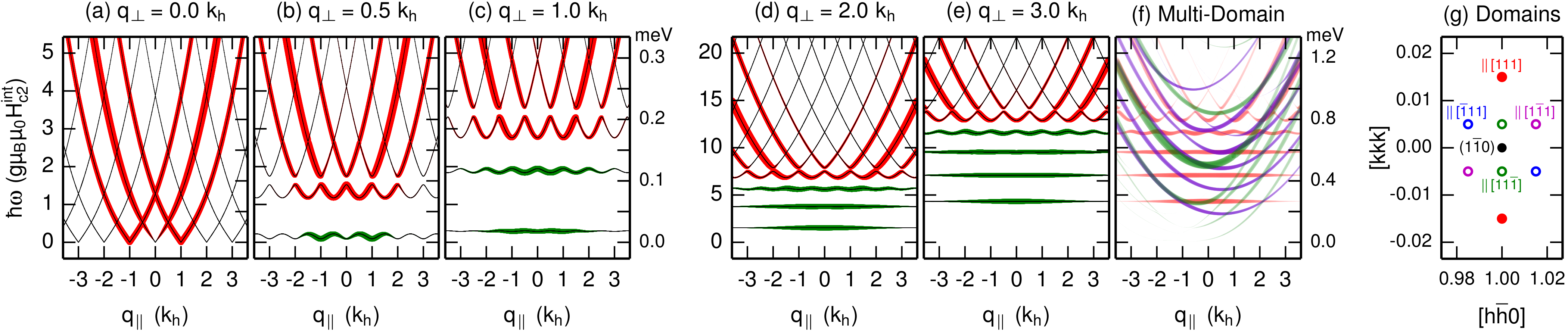}
\caption{(a)-(e) Universal helimagnon spectrum for momenta $q_\parallel$ along the pitch direction, ${\bf k}_h$, in the {\it repeated zone scheme}  (thin lines) at zero field and $\chi_{\rm con}^{\rm int} = 0.34$ for various values of perpendicular momentum $q_\perp$. 
The thickness of the colored lines indicate the theoretically expected intensity for inelastic neutron scattering on a domain with ${\bf k}_h \parallel [111]$ with respect to a nuclear $(1\bar10)$ reflection and for ${\bf q}_\perp \parallel [11\bar 2]$. 
The colors red and green distinguish between dispersive and practically flat helimagnon bands, respectively.
The energy units on the right axes correspond to MnSi at $20$ K for which $g \mu_B \mu_0 H_{\rm c2}^{\rm int} = 0.062$ meV.
%
%
%
(f) Weights of inelastic 
contributions arising from an equal population of all domains (color coding as in (g)) 
for $q_\perp = 3 k_h$ as a function of ${\bf q}_\parallel = q_\parallel \hat {\bf k}_h$ with respect to ${\bf k}_h \parallel [111]$ (red domain).
(g) Scattering plane around the nuclear $(1\bar10)$ peak with four domains that are degenerate at zero field; the open circles correspond to out-of plane pitch wavevectors projected onto the plane. 
}
\label{fig:1}
\end{figure*}

The spin-wave excitations of these magnetic crystals possess a band structure $\omega_{n,{\bf q}}$ with band index $n$ that, according to Bloch's theorem, reflects the periodicity of the magnetic order. For the magnetic helix with a pitch vector ${\bf k}_h$, the dispersion is periodic, $\omega_{n,\bf q} = \omega_{n,{\bf q} + m {\bf k}_h}$ with $m \in \mathds{Z}$, along the direction in momentum space singled out by ${\bf k}_h$. In contrast to commensurate antiferromagnets, however, the size $k_h = |{\bf k}_h|$ of the resulting magnetic Brillouin zone is small; for MnSi at lowest temperatures $\lambda_h = 180$\AA\, and $k_h = 0.035$\AA$^{-1}$. 
Importantly, this ensures on the one hand that the dispersion, $\omega_{n,{\bf q}}$, of the magnons is universal in the sense that it is captured by an effective continuum theory and is determined by only a few parameters.
On the other hand, a high resolution of momenta is required in order to resolve the band structure experimentally with the help of inelastic neutron scattering. In a first experiment on MnSi, Janoschek {\it et al.}~\cite{MJano:2010} succeeded to acquire scattering spectra for the helimagnons and also described them theoretically but neglecting dipolar interactions \cite{BelitzI,Petrova2011}.
In MnSi at zero magnetic field, the pitch vector of the helix, ${\bf k}_h$, aligns with one of the eight equivalent crystallographic $\langle 111 \rangle$ directions giving rise to four magnetic domains each of which is determined up to a phase. 
At that time, the experiment was performed on a single crystal containing multiple domains so that magnon excitation branches of all four domains were simultaneously excited. This resulted in broad total spectra rendering the identification of individual magnon modes impossible, see Fig.~\ref{fig:1}(f).  
Individual modes, however, have already been identified with the help of magnetic resonance experiments on MnSi, Cu$_2$OSeO$_3$, and Fe$_{0.8}$Co$_{0.2}$Si \cite{Date1977,onose,schwarze} that probe certain branches but only at the center of the Brillouin zone, i.e., at frequencies $\omega_{n,{\bf q = 0}}$. Schwarze {\it et al.} \cite{schwarze} were able to explain these resonances quantitatively after taking dipolar interactions into account.

In the present study, the band structure of the helimagnons could  be resolved by first preparing 
a crystal of MnSi in a single magnetic-helix domain by applying a small field. 
A finite magnetic field ${\bf H}$ competes with the crystalline anisotropies to align the pitch vector, ${\bf k}_h$, of the helix. For sufficiently large $H > H_{c1}$, the pitch vector is parallel to ${\bf H}$ as this allows the magnetization to cant towards the field gaining magnetic Zeeman energy giving rise to a single magnetic-helix domain. 
The magnetization ${\bf M} = m\, \hat n$ with $\hat n^2 = 1$ is then governed by the free energy density $\mathcal{F} = \mathcal{F}_{\rm mag}+\mathcal{F}_{\rm dipolar}$ where the first and second term account for the short-range magnetic and the dipolar interactions, respectively. In the low-energy limit the former reduces to $\mathcal{F}_{\rm mag} \approx \mathcal{F}_{0}$ with the standard model for chiral magnets \cite{Bak_Jensen}
\begin{align} \label{StandardM}
\mathcal{F}_0 = \frac{\rho_s}{2} \Big[ (\nabla_i \hat n_j)^2 + 2 k_{h0} \hat n (\nabla \times \hat n) \Big] 
- \mu_0 m \hat n {\bf H}.
\end{align}
Minimization of $\mathcal{F}_{0} + \mathcal{F}_{\rm dipolar}$ yields the conical helix $\hat n_0^T = (\sin \theta \cos (k_h z),\sin \theta \sin (k_h z), \cos \theta)$ for an applied field ${\bf H}$ along the $z$-axis with $k_h = k_{h0}$. 
The angle $\theta$ parametrizes the homogeneous part of the magnetization that depends linearly on the magnetic field $m \cos\theta = \chi^{\rm int}_{\rm con}  H_{\rm int}$ where the internal field $H_{\rm int} = H/(1 + N_z \chi^{\rm int}_{\rm con})$ with demagnetization factor $N_z$ arising from $\mathcal{F}_{\rm dipolar}$ and the susceptibility $\chi^{\rm int}_{\rm con} = \mu_0 m^2/(\rho_s k_h^2)$ being related to the stiffness density $\rho_s$. The transition to the field-polarized state then occurs at the second critical field 
$H^{\rm int}_{c2} = m/\chi^{\rm int}_{\rm con} = \rho_s k_h^2/(\mu_0 m)$.

Plugging the standard parametrization $\hat n = \hat e_3 \sqrt{1 - 2 |\psi|^2} + \psi \hat e^+ + \psi^* \hat e^-$ with $\hat e^\pm = (\hat e_1 \pm i \hat e_2)/\sqrt{2}$ and $\hat e_\alpha \hat e_\beta = \delta_{\alpha\beta}$
where $\hat e_3 \equiv \hat n_0$ into the Landau-Lifshitz equation and expanding in lowest order in $\psi$ one finds that for intermediate fields $H_{c1}<H<H_{c2}$, the magnon wave-function $\psi$ is governed by a bosonic Bogoliubov-deGennes equation $i \hbar \tau^z \partial_t \vec \Psi = \mathcal{H} \vec \Psi$ for the spinor $\vec \Psi^T = (\psi, \psi^*)$.
The Hamiltonian $\mathcal{H} = \mathcal{H}_{0}+\mathcal{H}_{\rm dipolar}$ consists of two parts that derive from $\mathcal{F}_0$ and $\mathcal{F}_{\rm dipolar}$, respectively. The former contribution reads \cite{MJano:2010,Petrova2011}
\begin{align} \label{Hamiltonian}
\mathcal{H}_{0}
&= \mathcal{D} \left[
-\mathds{1} \nabla^2
- i 2 \tau^z k_h \hat n_\perp({\bf r}) \nabla
+ \frac{k_h^2 \sin^2 \theta}{2}(\mathds{1} - \tau^x)\right]
\end{align}
where $\tau^x$ and $\tau^z$ are Pauli matrices, and the stiffness is $\mathcal{D} = g\mu_B \rho_s/m = 
g \mu_0 \mu_B H^{\rm int}_{c2}/k_h^2$ with $g\approx 2$ for MnSi.
The helimagnons are subject to an effective vector potential 
$\hat n^T_\perp({\bf r}) = (\sin \theta \cos (k_h z), \sin \theta \sin (k_h z),0)$ that is periodic and is responsible for the formation of bands. 

In the low-energy limit the Hamiltonian depends only on three parameters: $(i)$ the critical field $H^{\rm int}_{c2}$, $(ii)$ the pitch vector ${\bf k}_h$, and $(iii)$ the numerical value  $\chi^{\rm int}_{\rm con} = 0.34$  for MnSi that effectively measures the strength of the dipolar interactions \cite{schwarze}. All these parameters are known from independent measurements resulting in a universal prediction for the helimagnon spectrum that we discuss in the following.

For momenta ${\bf q} = q_\parallel \hat {\bf k}_h$ strictly longitudinal to the pitch vector, the helimagnons decouple from the periodic potential in Eq.~\eqref{Hamiltonian}. 
The helimagnon dispersion can then be obtained in closed form and reads in the {\it extended zone scheme}, see Fig.~\ref{fig:1}(a),
\begin{align} \label{SpectrumParallel}
&\hbar \omega({\bf q} = q_\parallel \hat {\bf k}_h)
= \mathcal{D} |q_\parallel| \sqrt{q_\parallel^2 + (1+\chi_{\rm con}^{\rm int})k_h^2 \sin^2 \theta}
\end{align}
where $\sin^2 \theta = 1 - (H_{\rm int}/H^{\rm int}_{c2})^2$ and the dependence on $\chi_{\rm con}^{\rm int}$ is attributed to the dipolar interaction. 
For large  momenta $|{\bf q}_\perp| \gg k_h$ perpendicular to the pitch, ${\bf q_\perp k}_h = 0$, on the other hand, the dipolar interaction as well as the last term in Eq.~\eqref{Hamiltonian} can be neglected and the wave equation reduces to a Schr\"odinger equation
\begin{align} \label{Mathieu}
&\hbar \omega \psi_{\omega,{\bf q}_\perp}(z) = 
\\\nonumber &\mathcal{D} \Big[q_\perp^2 - \partial_z^2 + 2 k_h q_\perp \sin\theta \cos(k_h z - \alpha)\Big]\psi_{\omega,{\bf q}_\perp}(z)
\end{align}
where $(q_x,q_y) = q_\perp (\cos \alpha,\sin\alpha)$. It effectively describes a particle in the presence of a periodic cosine potential along the $z$-axis, i.e., it is the {\it Mathieu equation}. Interestingly, the strength of the potential can be tuned by $q_\perp$ with the concomitant reconstruction of the spectrum as illustrated in Figs.~\ref{fig:1}(b)-(e). 
A finite $q_\perp$ activates Bragg reflections that open gaps at the Bragg planes.
As the potential only contains the two primary Fourier components, the size of these band gaps however strongly decreases as a function of the band index $n=0,1,2,...$ \cite{Avron1981}. For large $q_\perp \gg (n+1)^2 k_h/\sin\theta$ the potential for a given band $n$ eventually becomes strong and basically localizes the helimagnon along the pitch direction. In this tight-binding limit, the bands are practically flat, i.e., independent of $q_\parallel$, and the spectrum is given by
\begin{align}
 \hbar \omega_{n,{\bf q}} \approx \mathcal{D} \Big[& q_\perp^2 - 2 k_h q_\perp  \sin\theta
\\\nonumber&
 + 2 k_h \sqrt{k_h q_\perp  \sin\theta} \Big(n + \frac{1}{2}\Big) 
- \frac{k_h^2}{8}(n^2 + n) \Big],
\end{align}
up to corrections of order $\mathcal{O}(\mathcal{D} k_h^2)$. The second and third term derive from an expansion of the cosine potential up to quadratic order giving rise to a harmonic oscillator spectrum; the last term is attributed to the anharmonicity of this potential.

\begin{figure}[t]
\includegraphics[width=\linewidth] {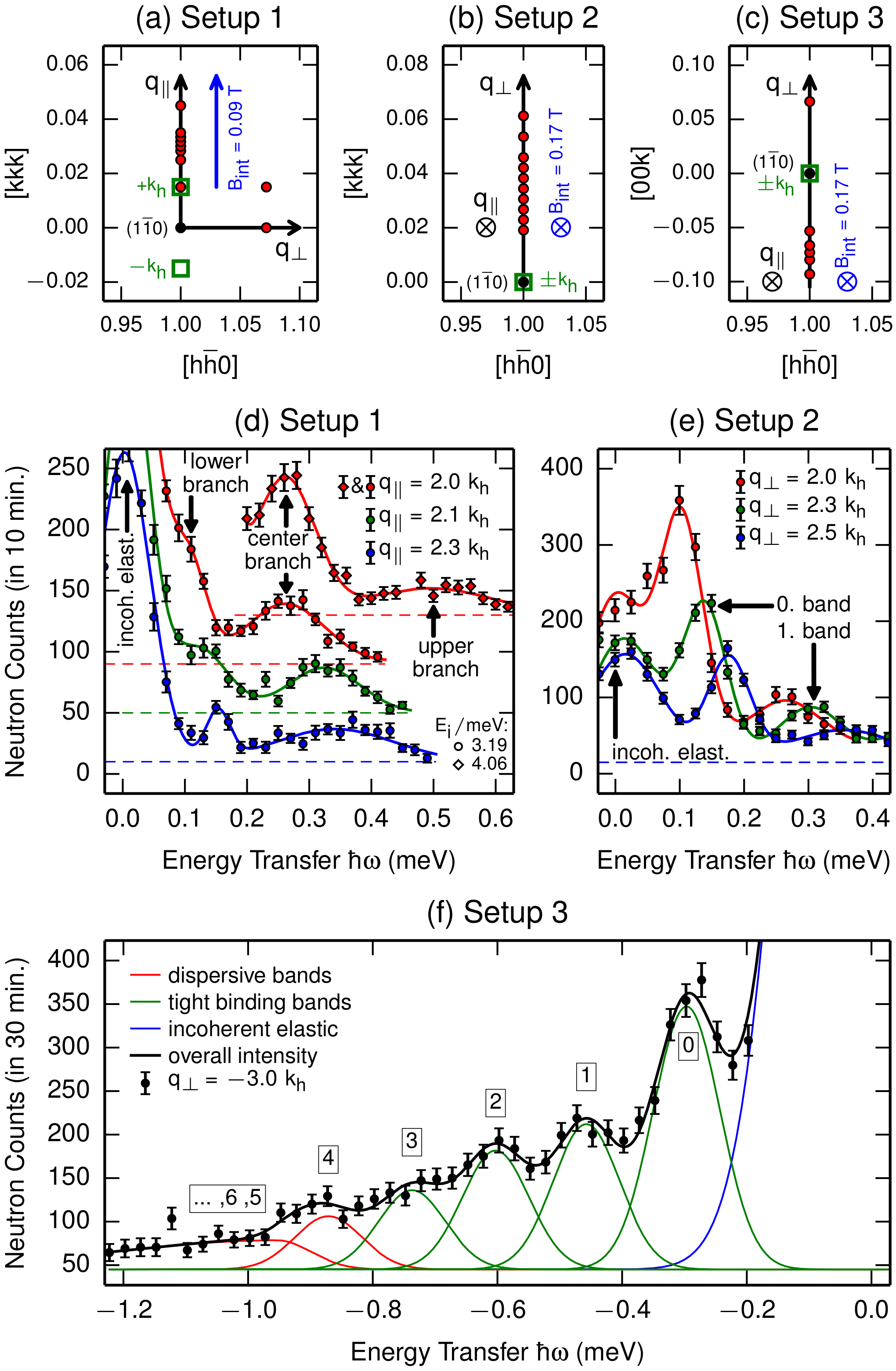}
\caption{
Panels (a)-(c) show the scattering planes for the setups 1, 2, and 3, respectively. 
The magnetic field (blue color code) defines the orientation of ${\bf k}_h$ resulting in magnetic satellite peaks (green squares); for panel (b) and (c) they are located above and below the scattering plane. Constant-$Q$ scans have been performed at the points in reciprocal space
marked by red circles. Panels (d)-(f) show examples of scans 
obtained with the three setups at 20~K, where the full lines are fits of 
multi-Gaussian profiles to the data. (d) Three helimagnon branches are measured with setup~1
(incident energies $E_i = 3.19$ and 4.06~meV), 
while (e) setup~2
($E_i = 4.06$~meV) and (f) setup~3 ($E_i = 5.04$~meV) clearly resolve the first 
two and five bands, respectively. The elastic peaks appear due to the 
$Q$-independent 
incoherent scattering. Note that the scans in (d) are shifted by 40 counts with respect to each other  for 
clarity.
}
\label{fig:2}
\end{figure}

In order to verify the helimagnon theory, we have conducted inelastic neutron scattering experiments at the cold neutron triple-axis spectrometer MIRA-2 \cite{Georgii:2015} at the neutron source FRM-II in Garching using 
neutrons with fixed incident energies $E_i = 3.19,\, 4.06$ and 5.04~meV. MIRA-2 
is particularly well suited for the experiments due to the intrinsic excellent 
momentum resolution, which is important for resolving spin waves close to  
magnetic satellite peaks. The collimations before and after the sample were 30' 
resulting in energy resolutions $\triangle E = 76,\, 118$ and 
161~$\mu$eV, respectively. 
Higher order neutrons were removed by a cooled Be-filter. 
Two large single crystals of MnSi were used in the present study with a volume of approximately 
8 cm$^3$ and a small mosaicity of 10'.

For the measurement of the helimagnon spectrum three experimental setups were realized, see Fig.~\ref{fig:2}(a)-(c).
A magnetic field $H > H_{c1}$ was applied in all setups to align the pitch vector and, more importantly, to prepare a single magnetic-helix domain. 
In setup 1, Fig.~\ref{fig:2}(a), the magnetic field $\mu_0 H_{\rm int} \approx 0.07$~T 
was within the scattering plane and pointing along the 
$[111]$ direction measured with respect to the nuclear $(1\bar10)$ Bragg peak. 
This configuration is ideal to obtain an excellent energy resolution for the 
dispersive branches along $q_\parallel$ with $q_\perp = 0$. In setup 
2, Fig.~\ref{fig:2}(b), a magnetic field $\mu_0 H_{\rm int} \approx 0.13$~T
was applied perpendicular to the scattering plane along $[11\bar{2}]$ aligning the pitch with $[11\bar{2}]$, and the helimagnons were measured for momenta ${\bf q}_\perp$ along $[111]$. In setup 3, Fig.~\ref{fig:2}(c), measurements were performed on the second crystal as a consistency check with a magnetic field $\mu_0 H_{\rm int} \approx 0.13$~T applied perpendicular to the scattering plane along the [110] direction orienting the pitch along [110], which allowed us to probe the bands for ${\bf q}_\perp$ along $[001]$. The data acquired along $[001]$ and $[111]$ turned out to be identical. 

\begin{figure*}[t]
\includegraphics[width=0.8\linewidth] {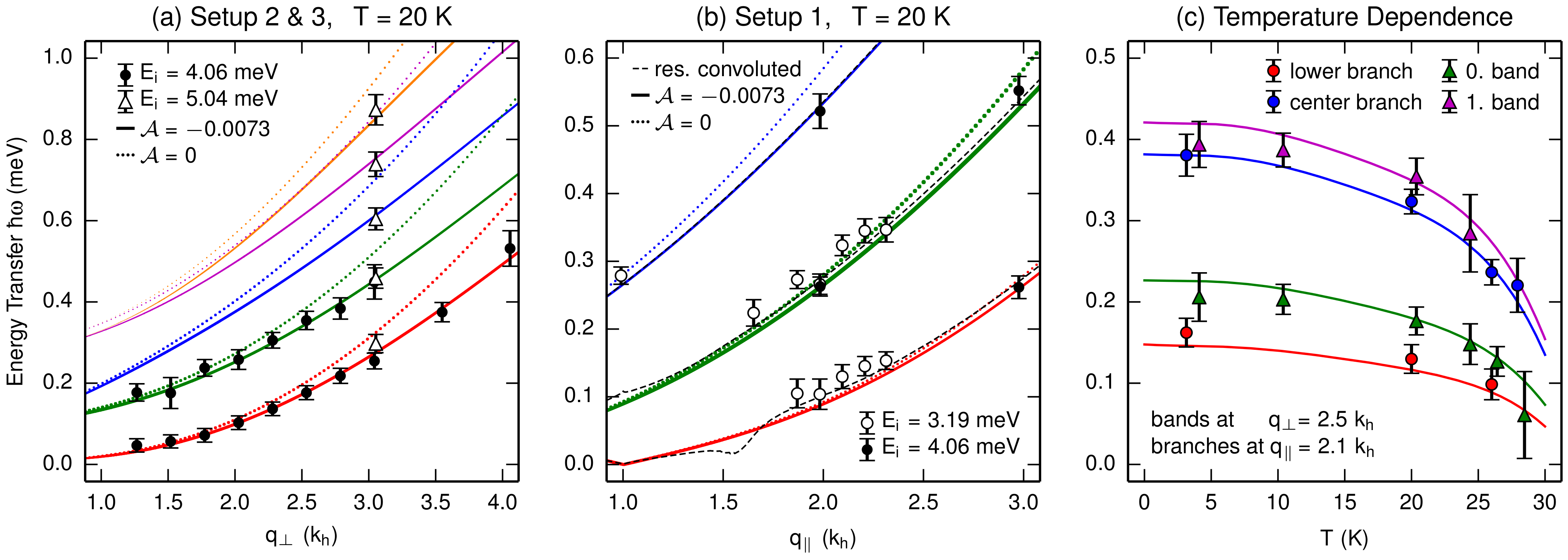}
\caption{Comparison between theory and experiment (symbols). (a) \& (b) Helimagnon spectrum as a function of $q_\perp$ and $q_\parallel$ with nominally $q_\parallel= 0$ and $q_\perp=0$, respectively. 
The universal spectrum (dotted lines) agrees with the data at low energies;
the correction due to Eq.~\eqref{Corr} is necessary (solid lines) for 
quantitative agreement at higher energies. 
The instrumental resolution slightly shifts the lower and center branch upwards
(black dashed line in panel (b)), see text. (c) $T$-dependence of the two helimagnon modes lowest in 
energy measured at certain fixed momenta.}
\label{fig:3}
\end{figure*}

Most measurements were performed at 20~K where the intensity of the magnetic 
Bragg peaks is high and the Bose factor is large yielding a high inelastic intensity. As an example, Fig.~\ref{fig:2}(d) shows a sequence of scans obtained with setup~1 with momenta longitudinal to ${\bf k}_h$, i.e., $q_\perp = 0$. 
According to Fig.~\ref{fig:1}(a) weights from three branches are expected and can indeed be identified for $q_\parallel = 2 k_h$. The high-energy mode near $\hbar \omega \approx 0.5$~meV is best seen for incident energy $E_i = 4.06$ meV ($\Diamond$ symbols) and the other two are resolved for $E_i = 3.19$ meV ($\ocircle$ symbols). While the low-energy mode at $\hbar \omega \approx 0.1$~meV is partly hidden by the incoherent scattering, the mode with intermediate energy at $\hbar \omega \approx 0.25$~meV is most prominent. Both shift to larger energies with increasing $q_\parallel$.
Fig.~\ref{fig:2}(e) shows the $q_\perp$-dependence of the first two 
bands which are accessible with setup 2 at $q_\parallel = 0$. With increasing $q_\perp$ the 
energy of the bands increases as expected. Finally using setup 3, we succeeded to probe the annihilation of helimagnons that belong to at least five different bands for ${\bf q}_\perp = - 3 k_h (0,0,1)$, see Fig.~\ref{fig:2}(f).

For a quantitative analysis, the data was fitted to a multi-Gaussian spectrum 
after subtracting a background of 1.0 cts/min and 1.5 cts/min for setup~1 
and setups~2~\&~3, respectively. The latter was determined by 
repeating most measurements at 3 K, 
where the inelastic magnetic scattering is weak. 
The fits are shown as solid lines in Figs.~\ref{fig:2}(d)-(f). 
Note that the observed width varies for different peaks because the resolution of the spectrometer depends on the energy transfer $\hbar\omega$. The peak positions obtained from these fits are summarized in Fig.~\ref{fig:3}.

Fig.~\ref{fig:3}(a) and (b) compares the experimental data 
at $20$~K to the prediction of the universal helimagnon spectrum (dotted lines) using 
$k_h \approx 0.036$\AA$^{-1}$ \cite{Grigoriev} and $\mu_0 H^{\rm int}_{c2} \approx 0.53$ T \cite{Bauer2010}, 
that determines the energy scale $g\mu_B\mu_0 H^{\rm int}_{c2} = 0.062$ meV at $20$ K.
While there is perfect 
agreement at low energies, deviations become substantial at higher energies with 
the data assuming systematically lower values than predicted. It turns out that 
these deviations can be attributed to corrections to the low-energy theory 
\eqref{StandardM} caused by higher-order contributions in the gradient 
expansion. Representatively, we consider the following correction \cite{Remark}
\begin{align} \label{Corr}
\delta \mathcal{F}_{\rm mag} = \frac{\rho_s}{2} \frac{ \mathcal{A}}{k_{h0}^2} (\nabla^2 \hat n)^2
\end{align}
where $\mathcal{A}$ is a dimensionless number that is expected to be small and of order $\mathcal{A} \sim \mathcal{O}((a k_{h})^2)$ with $a^3 \approx 24$ \AA$^3$ 
being the volume of a formula unit in MnSi so that $(a k_{h})^2 \approx 0.01$. For 
momenta $q \approx 3 k_{h}$, for example, the corrections arising from Eq.~\eqref{Corr} are expected to be 
$3^2 (a k_{h})^2 \sim 10\%$ as compared to the first term in Eq.~\eqref{StandardM}, 
which is in fact comparable to the deviations observed in Fig.~\ref{fig:3}. 
For a quantitative comparison in the entire investigated energy range, we took 
the modification of the helimagnon spectrum due to Eq.~\eqref{Corr} into 
account providing us with a single fit parameter $\mathcal{A}$. The solid lines 
in Fig.~\ref{fig:3}(a)-(c) show a best-fit 
yielding $\mathcal{A}_{\rm fit} = - 0.0073 \pm 0.0004$.

The bands in Fig.~\ref{fig:3}(a) are mostly in the tight-binding limit and thus practically independent of $q_\parallel$. As a result, the measured peak positions are rather insensitive to the vertical momentum resolution of MIRA  ($\Delta q \approx \pm 0.6 k_h$). 
The flatness of the lowest bands has been also explicitly checked by comparing the two scans at finite $q_\perp$ in Fig.~\ref{fig:2}(a) using setup~1 (not shown).
In contrast, the dispersive bands in Fig.~\ref{fig:3}(b) have a substantial $q_\perp$ dependence which is sampled by the instrumental momentum resolution. The resulting admixture of spectral weight with finite $q_\perp$ into the nominal $q_\perp = 0$ spectrum leads to a slight upward shift of the expected peak positions as indicated by the black dashed line in panel Fig.~\ref{fig:3}(b) \cite{Remark2}. 


In order to investigate the temperature dependence, the helimagnons were 
measured at $q_\parallel = 2.1 k_h$ with setup 1 and at $q_\perp = 2.5 k_h$ with 
setup 2 at four and six different temperatures, respectively. 
Fig.~\ref{fig:3}(c) shows the $T$-dependence of the peak positions for the two 
helimagnon modes with lowest energies. 
The $T$-dependence mainly derives from 
the magnitude of the local magnetization $m(T)$ that is reflected in the 
critical field $H^{\rm int}_{c2}(T)$ entering the stiffness $\mathcal{D}$ in 
Eq.~\eqref{Hamiltonian}. 
The theoretically expected peak positions resulting 
from the values of $H^{\rm int}_{c2}(T)$ obtained from ac susceptibility 
measurements \cite{Bauer2010} as well as from the weakly $T$-dependent pitch length $k_h(T)$ \cite{Grigoriev}
are shown as solid lines in Fig.~\ref{fig:3}(c) in good agreement with experiment.  


In the present work, we experimentally verified the theory for helimagnons and established the emergence of flat helimagnon bands at finite momenta. The latter might be technologically exploited, for example, in the design of magnon waveguides in magnonic applications \cite{Krawczyk2014}. 
Our findings set the stage for further investigations of helimagnons at finite magnetic field and pressure
and, in particular, 
their influence on 
the stabilization of various phases \cite{muehlbauer2009,Brazowskii,Bauer2013,Buhrandt2013}. 
It has been suggested that the renormalizations due to magnons are crucial for the stabilization of the 
skyrmion crystal \cite{muehlbauer2009}. They might be also important for the formation of the enigmatic non-Fermi liquid phase \cite{Kirkpatrick2010,Watanabe2014} observed in MnSi at high pressures \cite{Pfleiderer2001,RRitz:2013}, 
a long-standing puzzle of this material that still remains unsolved.


We thank Reinhard Schwikowski, Andreas Mantwill, and the machine shop of the 
FRM II for their technical support. We thank Sarah Dunsinger for donating a 
sample holder and Tobias Weber for the extensive IT-support. This work was 
supported by the DFG under GE971/5-1 and TRR 80, and by ERC grant 291079 (TOPFIT). M.~J.~was funded by the Los Alamos National Laboratory Directed Research and Development program.

\bibliography{paper}

\begin{thebibliography}{28}
\bibitem[{Rem({\natexlab{a}})}]{Remark}
\bibinfo{note}{There are further corrections of similar importance, e.g.,
  $\delta \mathcal{F}_{\rm mag} \sim \hat n (\partial_x^4 +
  \partial_y^4+\partial_z^4)\hat n$. The available data cannot discriminate
  between the various corrections so that we limit ourselves to Eq.~(6).}
\bibitem[{Rem({\natexlab{b}})}]{Remark2}
\bibinfo{note}{The dip close to $q_\parallel \approx 1.5 k_h$ is attributed to
  Bragg reflections that open a gap for finite $q_\perp$ between the lowest two
  bands, cf.~Fig.~1(b).}
\end{thebibliography}

\end{document}